\documentclass{emulateapj}
\submitted{ApJ, in press} 

\newcommand{\snr}{G315.9$-$0.0}
\newcommand{\pwn}{G315.78$-$0.23}
\newcommand{\psr}{PSR~J1437$-$5959}
\newcommand{\kms}{\,km\,s$^{-1}$}
\newcommand{\ergs}{\,erg\,s$^{-1}$}

\newcommand{\mjb}{\,mJy\,beam$^{-1}$}
\newcommand{\edot}{\mbox{$\dot{E}$}}

\begin{document}
\title{AN EXTREME PULSAR TAIL PROTRUDING FROM THE FRYING PAN SUPERNOVA REMNANT}

\shorttitle{An Extreme PWN in \snr}
\shortauthors{Ng et al.}

\author{C.-Y.\ Ng\altaffilmark{1,2,7,8}, N.\ Bucciantini\altaffilmark{3,4},
B.\ M.\ Gaensler\altaffilmark{2}, F.\ Camilo\altaffilmark{5},
S.\ Chatterjee\altaffilmark{6}, and A.\ Bouchard\altaffilmark{1}}
\altaffiltext{1}{Department of Physics, McGill University, Montreal, QC H3A
2T8, Canada}
\altaffiltext{2}{Sydney Institute for Astronomy, School of Physics, The University of Sydney, NSW 2006, Australia}
\altaffiltext{3}{NORDITA, Albanova Research Center, Roslagetullsbacken 23, 106 91 Stockholm, Sweden}
\altaffiltext{4}{INAF -- Osservatorio Astrofisico di Arcetri, Largo E. Fermi 5, 50125 Firenze, Italy}
\altaffiltext{5}{Columbia Astrophysics Laboratory, Columbia University, New York, NY 10027, USA}
\altaffiltext{6}{Astronomy Department, Cornell University, Ithaca, NY 14853,
USA}
\altaffiltext{7}{Tomlinson Postdoctoral Fellow}
\altaffiltext{8}{CRAQ Postdoctoral Fellow}

\email{ncy@physics.mcgill.ca}

\begin{abstract}
The Frying Pan (\snr) is a radio supernova remnant with a peculiar
linear feature (\pwn) extending 10\arcmin\ radially outward from the
rim of the shell. We present radio imaging and polarization observations
obtained from the Molonglo Observatory Synthesis Telescope and the Australia
Telescope Compact Array, confirming \pwn\ as a bow-shock pulsar wind nebula
(PWN) powered by the young pulsar J1437$-$5959. This is one of the longest pulsar
tails observed in radio and it has a physical extent over 20\,pc. We found a
bow-shock stand-off distance of 0.002\,pc, smallest among similar systems,
suggesting a large pulsar velocity over 1000\kms\ and a high Mach number
$\sim$200. The magnetic field geometry inferred from radio polarimetry shows a
good alignment with the tail orientation, which could be a result of high flow
speed. There are also hints that the postshock wind has a low magnetization
and is dominated by electrons and positrons in energy. This study shows that
PWNe can offer a powerful probe of their local environment,
particularly for the case of a bow shock where the parent supernova shell is
also detected.
\end{abstract}


\section{INTRODUCTION}

Neutron stars are born with large space velocities of a few hundred kilometers
per second. They eventually escape their natal supernova remnants (SNRs) after
$\sim10^5$\,yr and travel supersonically through the interstellar medium
(ISM), which typically has a sound speed of 1--10\kms\ in the cold and warm
phases. In this case, the pulsars' relativistic outflows are confined by ram
pressure of the ISM, resulting in bow-shock pulsar wind nebulae (PWNe). These
systems exhibit cometary morphologies with long tails, and emit broadband
synchrotron radiation from radio to X-rays \citep[see,][for a review]{gs06}.
As compared to PWNe within SNRs, bow-shocks are governed by simpler boundary
conditions, therefore, providing a powerful probe of the local ISM environment.

At radio frequencies, PWNe generally show a larger extent than their X-ray
counterparts, due to long synchrotron cooling timescale of radio-emitting
particles, which can be up to $10^5$--$10^6$\,yr. Therefore, radio PWNe can
act as direct calorimeters to reflect the integrated history of the systems.
Table~\ref{table:pwn} lists the longest pulsar tails detected in radio. The
best example is the Mouse PWN, its morphology consists of a broad cone-shaped
outer region and a collimated narrow inner region, which were suggested to be
postshock flows from the forward and backward termination shocks, respectively
\citep{gvc+04}. Radio observations of PWNe also reveal the nebular magnetic
field structure via polarimetry. Previous studies indicate a diverse $B$-field
configuration among bow-shock PWNe, from a helical field perpendicular to the
tail \citep{ngc+10} to a $B$-field parallel to the tail \citep{yg05}, but the
origin of the variation remains a puzzle.

The ``Frying Pan'' SNR (\object{\snr}) is a faint 15\arcmin--diameter radio
shell with an intriguing 10\arcmin\ linear protrusion, \object{\pwn} (the
``handle''), extending northwest from the rim, such that the overall
morphology resembles a Frying Pan (see Figure~\ref{fig:img}). Based on the
lack of an IR counterpart, \citet{wg96} concluded that the radio emission is
nonthermal and the source is an SNR. In this case, we suggested that the highly
unusual linear feature is due to the trailing relativistic wind of a fast
pulsar moving outward from the central birth place. We have recently performed
a deep radio search near the tip of the protrusion and discovered a new radio
\object[PSR J1437-5959]{pulsar, J1437$-$5959}, providing strong support to the
above picture \citep{cng+09}. This pulsar has a short period of 61.7\,ms, a
characteristic age of 110\,kyr, and a high spin-down power \edot=$1.5\times
10^{36}$\ergs. While its large dispersion measure of 549\,pc\,cm$^{-3}$
suggests a distance of 8\,kpc, as \citet{cng+09} noted, the estimate is
uncertain and could be easily off by 25\%.

\begin{deluxetable*}{lccccccc}
\tablecaption{Long Pulsar Tails in Radio, Sorted in Physical Length\label{table:pwn}}
\tablewidth{0pt}
\tabletypesize{\small}
\tablehead{\colhead{PWN}& \colhead{SNR} & \colhead{PSR/Neutron Star}&
\colhead{Length}&\colhead{Velocity}& \colhead{Distance\tablenotemark{1}}& 
\colhead{Reference}\\
& & & \colhead{(pc) }& \colhead{(\kms) }& \colhead{(kpc) }& }
\startdata
\pwn& Frying Pan & J1437$-$5959 & 20 & $\sim$1000& 8 & This work\\
Mouse &\nodata & J1747$-$2958 & 16 & 300 & 5 & 1,2 \\
G319.9$-$0.7 &\nodata & J1509$-$5850 &10 & 300 & 3 & 3,4 \\
G68.77+2.82 & CTB 80 & B1951+32 & 5 & 274 & 2 & 5,6 \\
G34.56$-$0.50 & W44 & B1853+01 & 2 & 375 & 2.6 & 7\\
Duck &\nodata& B1757$-$24 & 1.5 & $<$360 & 5.2 & 6 \\
G47.38$-$3.88 &\nodata& B1929+10 & 1.3 & 177 & 0.36 &8,9 \\ 
G189.23+2.90 & IC 443 & J061705.3+222127 & 1 & 230 & 1.5& 10,11 \\
N157B\tablenotemark{2} & N157B & J0537$-$6910 & 30 & ? & 50 & 12\\
G309.92$-$2.51\tablenotemark{3} &\nodata& J1357$-$6429 & 7 & ? & 2.5 & 13
\enddata
\tablenotetext{1}{All distance estimates are subject to large uncertainties,
except for PSR B1929+10, which is obtained from parallax.}
\tablenotetext{2}{This system is more likely to be a PWN crushed by the SNR
reverse shock, instead of a bow-shock \citep{van04,cwg+06}.}
\tablenotetext{3}{The radio counterpart is not confirmed.\vspace*{2mm} }
\tablerefs{(1) \citet{gvc+04}; (2) \citet{hgc+09}; (3) \citet{kmp+08}; (4)
\citet{ngc+10}; (5) \citet{cdg+03}; (6) \citet{zbc+08}; (7) \citet{fgg+96};
 (8) \citet{ccv+04}; (9) \citet{bkj+06};  (10) \citet{ocw+01};
(11) \citet{gcs+06}; (12) \citet{ldh+00}; (13) \citet{cpk+11}.}
\end{deluxetable*}

In this paper, we focus on the handle of the Frying Pan and establish its PWN
nature using new and archival radio observations from the Molonglo Observatory
Synthesis Telescope (MOST) and the Australia Telescope Compact Array (ATCA).
Its projected length over 20\,pc makes it one of the longest pulsar tails
observed. We show that this unique object can provide crucial information on
the pulsar environment that is not easily accessible in other systems. 

\section{OBSERVATIONS AND DATA REDUCTION}
\label{s2}
We carried out new radio imaging observations of the Frying Pan with the ATCA
at 3 and 6\,cm in four different array configurations, and processed archival
ATCA data at 13 and 20\,cm, including those from the Southern Galactic Plane
Survey \citep{hgm+06}. Table~\ref{table:obs} lists the parameters of all
datasets used in this study. The last two observations in the list were
carried out after the Compact Array Broadband Backend (CABB) upgrade
\citep{wfa+11}, which provides 2\,GHz bandwidth, nearly 20 times improvement
over the pre-CABB era. Note that with such a large bandwidth, the 3 and 6\,cm
data and images we refer to throughout this paper actually cover 3.0--3.7\,cm
and 4.7--6.6\,cm, respectively.
The pre-CABB observations at 3 and 6\,cm were made with a two-pointing mosaic to
cover 5\arcmin\ from the tip of \snr. One more pointing was added in the CABB
observations to extend the coverage over the entire 8\arcmin-handle. The flux
density scale is set by observations of the primary calibrator, PKS
B1934$-$638. A secondary calibrator, PKS B1414$-$59, which is 2\fdg4 away from
our source, was observed every 30\,minutes to determine the antenna gains.

\begin{deluxetable*}{lccccccc}
\tablecaption{Parameters for the ATCA Observations of \snr \label{table:obs}}
\tablewidth{0.8\textwidth}
\tabletypesize{\small}
\tablehead{\colhead{Obs. Date} & \colhead{Array}& \colhead{Center Frequency} &
\colhead{Usable Bandwidth} & \colhead{No. of} & \colhead{Integration} \\
& \colhead{Configuration} & \colhead{(MHz)} &
\colhead{(MHz)\tablenotemark{1}} & \colhead{Channels\tablenotemark{1}}
& \colhead{Time (hr)}}
\startdata
1998 Jul 5 & 750E & 1384, 2496 & 104 & 13 & 11 \\ 
1998 Dec -- 2000 Aug\tablenotemark{2} & various\tablenotemark{2}
 & 1384 & 104 & 13 & 0.33 \\
1999 Mar 25 & 1.5B & 1384, 2496 & 104 & 13 & 12 \\ 
1999 Apr 20 & 1.5C & 1384, 2496 & 104 & 13 & 12 \\ 
2008 Jul 23 & H214 & 1344, 1436 & 104 & 13 & 9 \\ 
2008 Dec 8 & 750B & 4800, 8640 & 104 & 13 & 11.5\\
2009 Feb 18 & EW352 & 4800, 8640 & 104 & 13 & 12\\
2009 Jun 13 & 6A & 5500, 9000 & 1848 & 1848 & 12\\
2009 Jul 31 & 1.5A & 5500, 9000 & 1848 & 1848 & 12.5 
\enddata
\tablenotetext{1}{Per center frequency.}
\tablenotetext{2}{Part of the Southern Galactic Plane Survey \citep{hgm+06}.
\vspace*{2mm}}
\end{deluxetable*}

All data reduction was performed using the MIRIAD
package\footnote{\url{http://www.atnf.csiro.au/computing/software/miriad/}}.
After discarding the edge channels and channels affected by self-interference,
we obtained a usable bandwidth of 104\,MHz and 1848\,MHz for the pre-CABB and
CABB data, respectively, at each wavelength. The longest baselines in the 13
and 20\,cm data were excluded to provide a uniform \emph{u-v} coverage. We
examined the visibility data to reject bad data points due to poor atmospheric
phase stability, then applied gain, bandpass, flux, and polarization
calibrations. Mosaiced intensity maps were formed using the multifrequency
synthesis technique with uniform weighting at 13 and 20\,cm and with natural
weighting at 3 and 6\,cm. The maps are deconvolved using a maximum entropy
algorithm \citep[\texttt{MOSMEM;}][]{sbd99} and restored with Gaussian beams
of FWHM $23\arcsec \times 20\arcsec$ at 20\,cm, $12\arcsec \times 9\farcs7$ at
13\,cm, and $3\farcs7 \times 3\farcs3$ at both 3 and 6\,cm. The final maps
have rms noise of 1, 0.1, 0.011, and 0.016\,mJy\,beam$^{-1}$ at 20, 13, 6, and
3\,cm, respectively. While the last two values are very close to the
theoretical limits, the 20\,cm map is severely contaminated by a strong source
nearby and the noise in the 13\,cm map is also slightly higher than the
theoretical level of 0.06\,mJy\,beam$^{-1}$.

For polarimetry, we only focused on the 3 and 6\,cm data since the other
observations lack the sensitivity and suffer from beam depolarization. To
mitigate the problem of bandwidth depolarization due to the large fractional
bandwidth of CABB data, we divided the 6\,cm CABB data into 33 sub-bands
(56\,MHz each) and the pre-CABB data into 2 sub-bands. The problem is less
severe at 3\,cm, therefore, we only divided the CABB data into 5 sub-bands.
Individual Stokes Q and U maps for the sub-bands are formed using the same
procedure as above for Stokes I. We attempted to apply the rotation measure
(RM) synthesis technique \citep{bd05} to determine the foreground Faraday
rotation. Although the result is generally consistent with the pulsar RM
reported by \citet{cng+09}, the uncertainty is too large to be useful.
Therefore, we adopted the pulsar value of $-700\pm25$\,rad\,m$^{-2}$ to
derotate the individual Stokes Q and U maps. The error in RM corresponds to
uncertainties of 1\fdg5 and 4\fdg3 in the polarization angles at 3 and 6\,cm,
respectively. Finally, the derotated maps in each frequency channel were
weighted according to their rms noise and co-added to generate maps of
polarized intensity and position angle.

In addition to the radio observations, we also processed the \emph{Fermi}
Large Area Telescope data to search for pulsations in $\gamma$-rays. We
carried out the analysis using the Fermi Science Tools v9r18p6 and considered
only the ``diffuse'' class of events with a cut at zenith angles
$>$105\arcdeg. We only used the first year of \emph{Fermi} data taken before
2009 September 27, when the pulsar ephemeris given by \citet{cng+09} is valid. We
extracted 0.1--10\,GeV photons within 1\arcdeg\ of the pulsar position,
applied the barycentric correction to their arrival times, and folded them
according to the radio ephemeris. However, no pulsation is detected. This is
not unexpected given the large pulsar distance. 

\section{RESULTS}

\subsection{Overall Morphology and Flux Densities}
\begin{figure*}[th]
\epsscale{1.1}
\plottwo{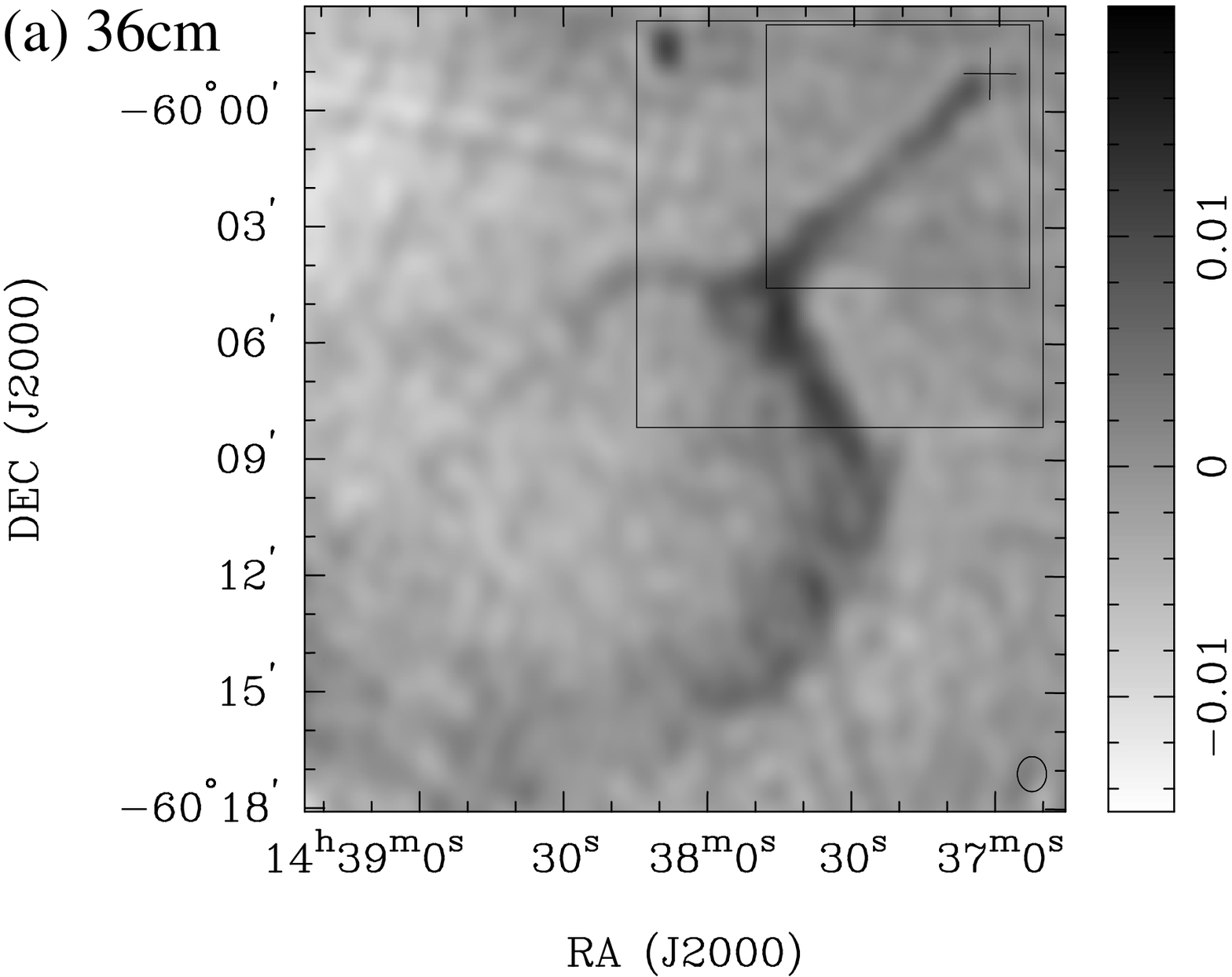}{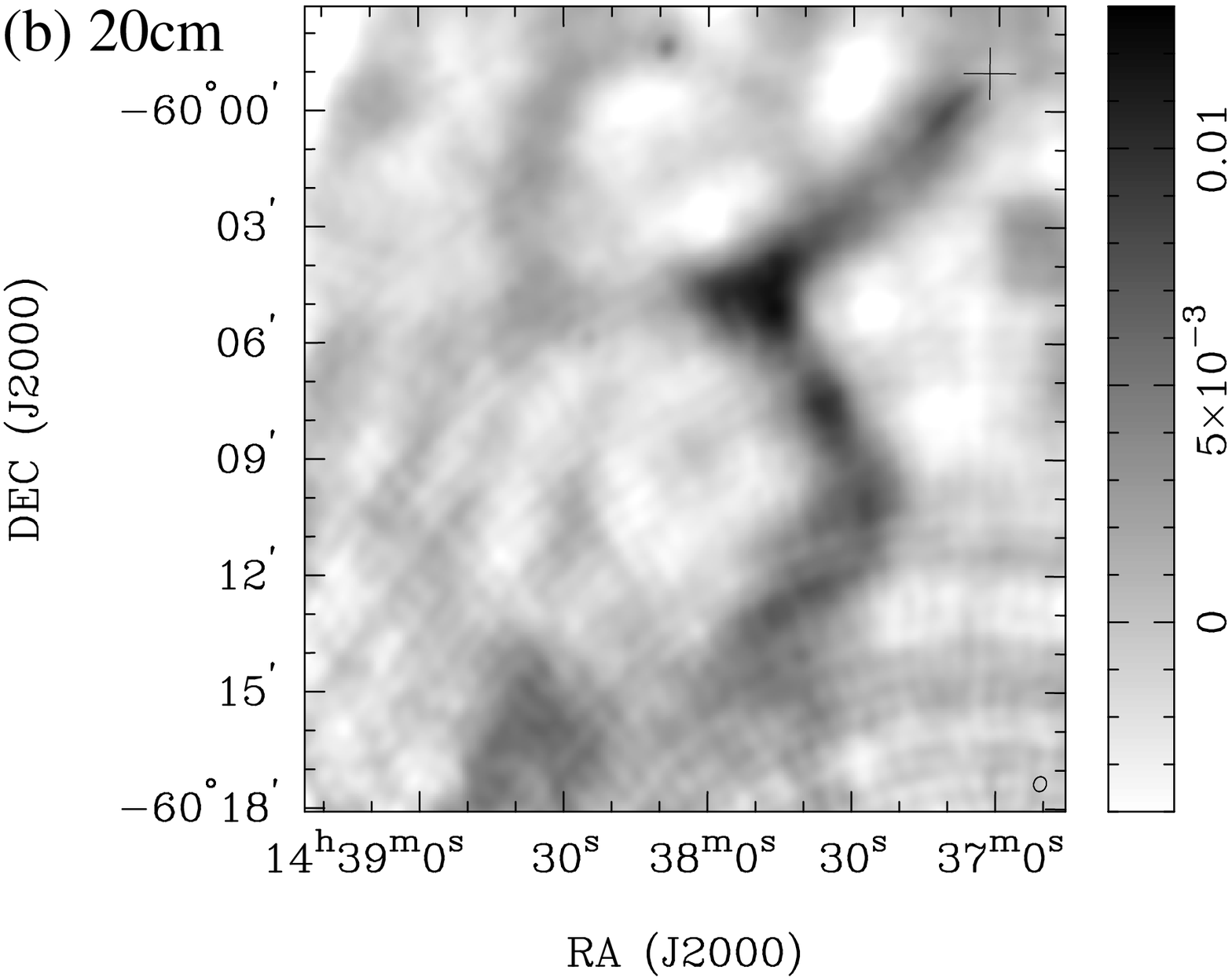}
\plottwo{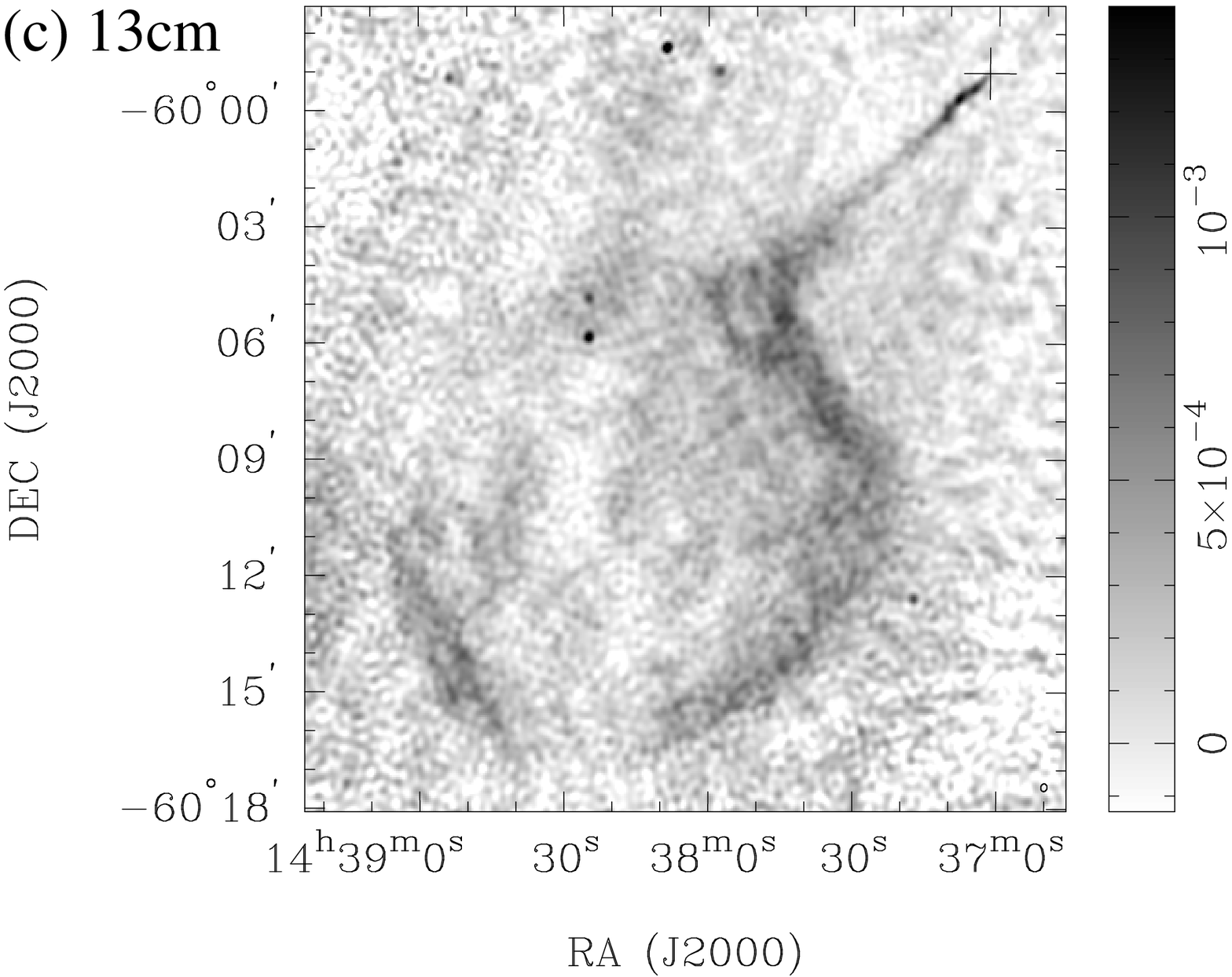}{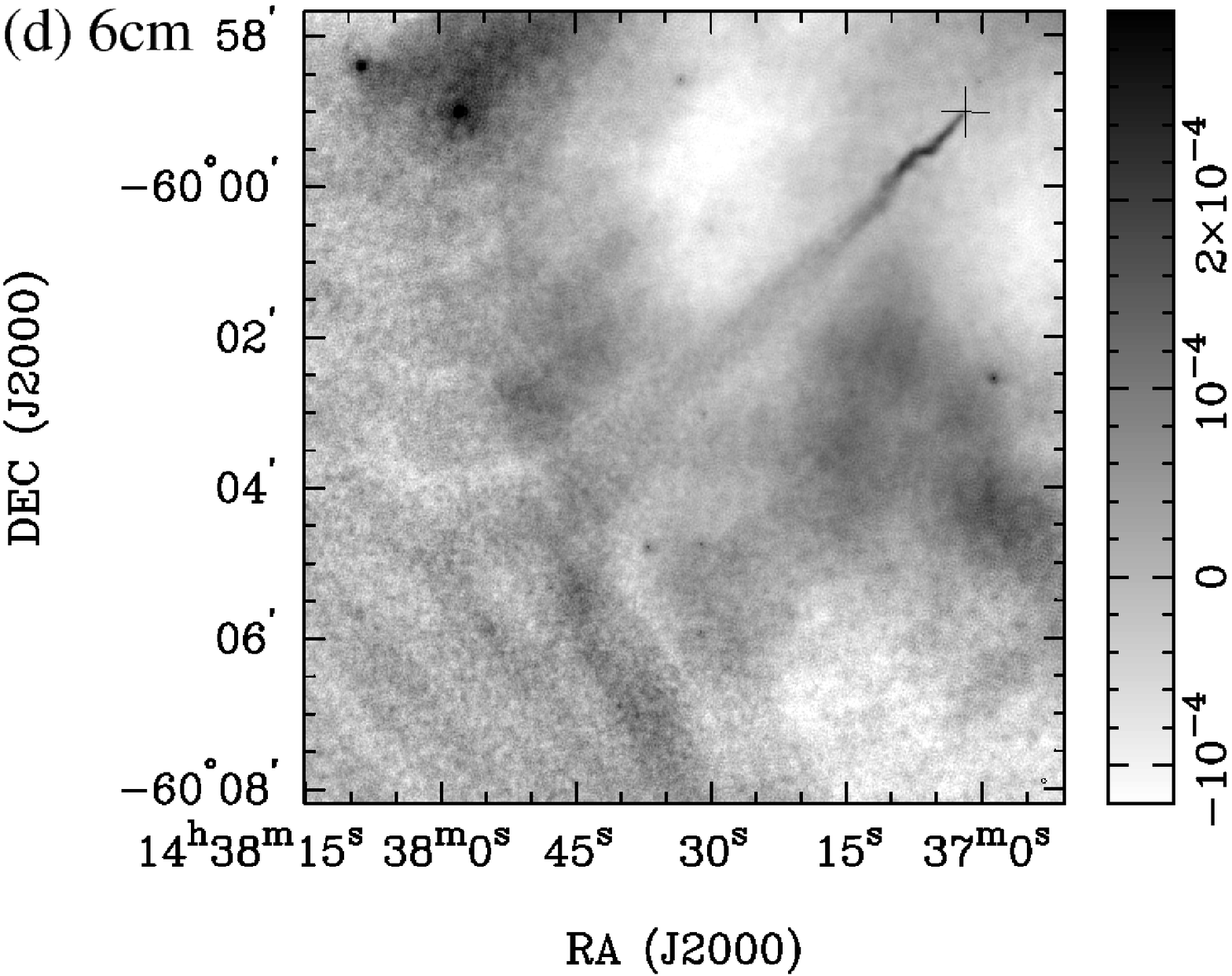}
\begin{minipage}[l]{0.5\textwidth}
\hspace*{-3mm}
\plotone{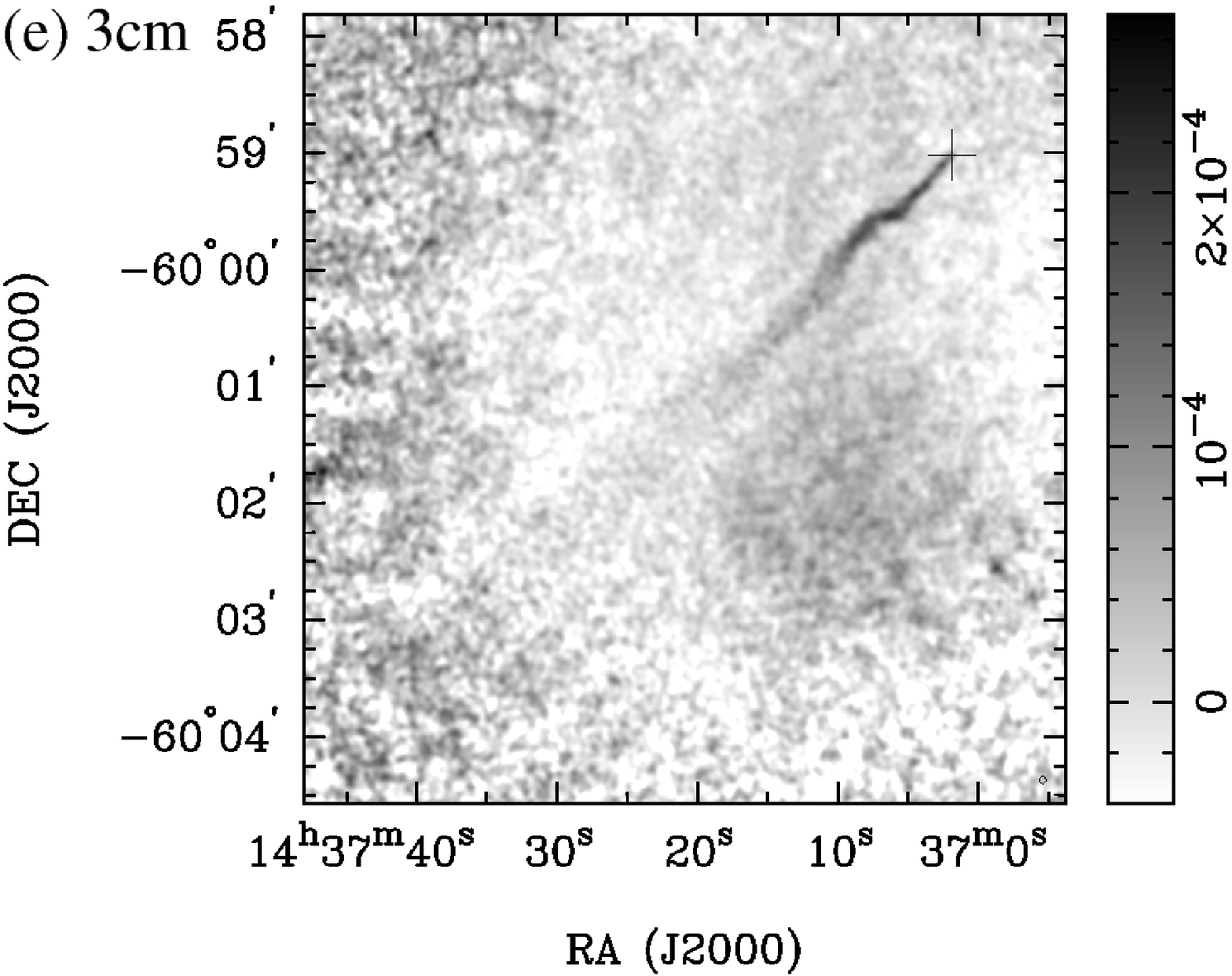}
\end{minipage}
\begin{minipage}[c]{0.45\textwidth}
   \hfill
 \end{minipage}
\caption{Radio continuum images of \snr\ at different wavelengths. The 36\,cm
image is adopted from the MOST Supernova Remnant Catalogue \citep{wg96} and
the rest are taken with ATCA. The large and small boxes in the 36\,cm image
indicate the field of views of the 6 and 3\,cm images, respectively. Position
of \psr\ [(J2000.0) R.A.=$\mathrm{14^h37^m01\fs91(3)}$,
decl.=$-59\arcdeg59\arcmin 01\farcs4(3)$ \citep{cng+09}] is marked by the
cross. The restoring beams are shown at the lower right of each panel. The
gray scales are linear with the scale bars in units of Jy\,beam$^{-1}$. The
13\,cm map is same as the one reported in \citet{cng+09}. \label{fig:img} }
\end{figure*}

Figure~\ref{fig:img} presents radio continuum images of the Frying Pan at 36,
20, 13, 6 and 3\,cm. All of them were taken with the ATCA as described in
Section \ref{s2} except the 36\,cm one, which is from the MOST Supernova Remnant
Catalogue\footnote{\url{http://www.physics.usyd.edu.au/sifa/Main/MSC}}
\citep{wg96}. The 13\,cm image has been reported in our previous paper
\citep{cng+09}. The radio shell exhibits a circular morphology with inner and
outer radii of 4\farcm5 and 7\arcmin, respectively. The overall shell is
brighter to the west, particularly in the northwest where it intersects with
the pulsar tail, but it is very faint and diffused to the northeast. Excluding
the handle, the shell has flux densities of $0.4\pm0.1$ and $0.5\pm0.1$\,Jy at
36 and 13\,cm, respectively. The image at 20\,cm is severely contaminated by
strong sources nearby, precluding any detailed flux measurement. Also, the 3
and 6\,cm observations were designed to map the tip of the linear protrusion,
hence, they are insensitive to the overall SNR structure.

\begin{figure*}[ht]
\epsscale{1.08}
\plottwo{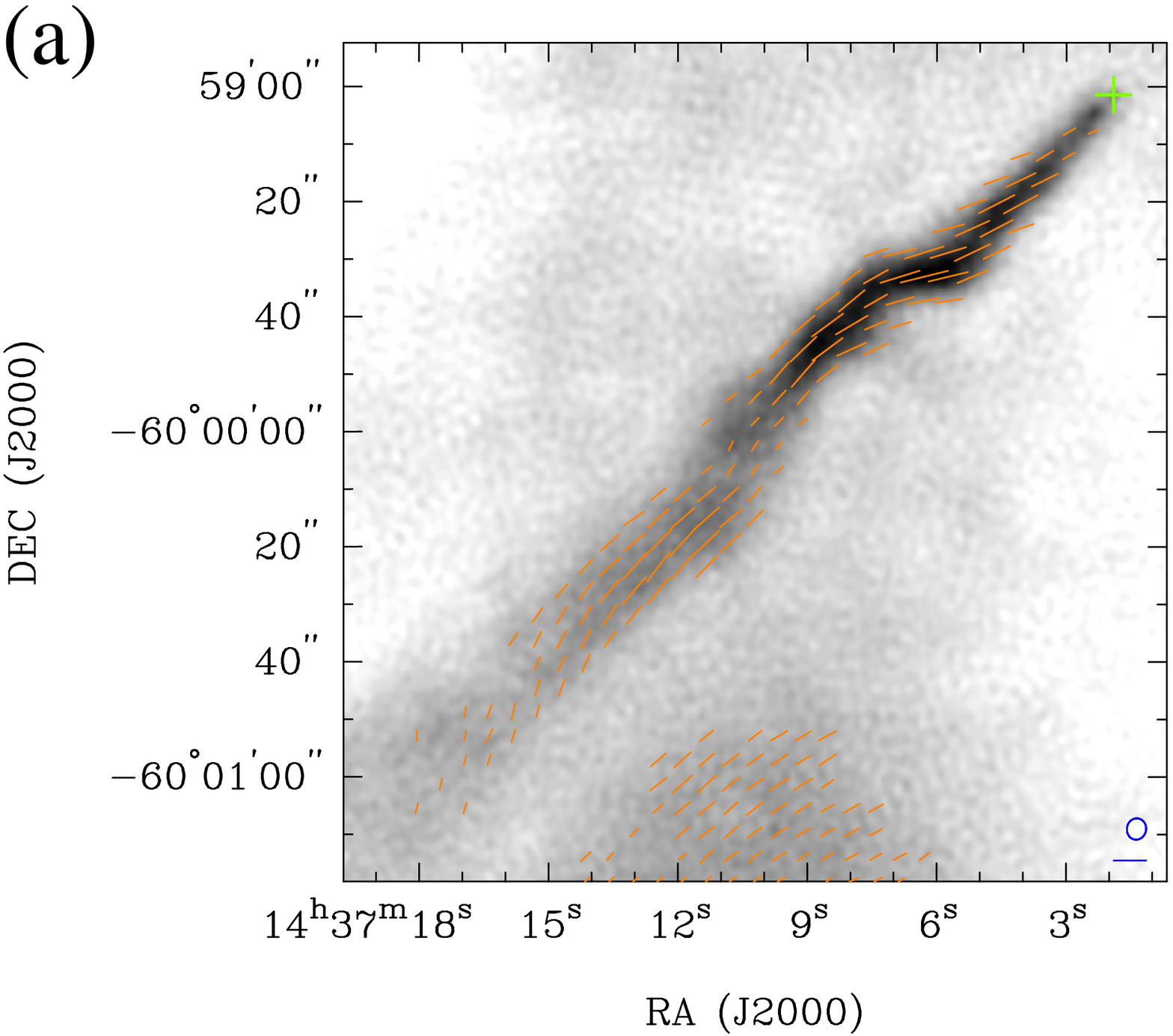}{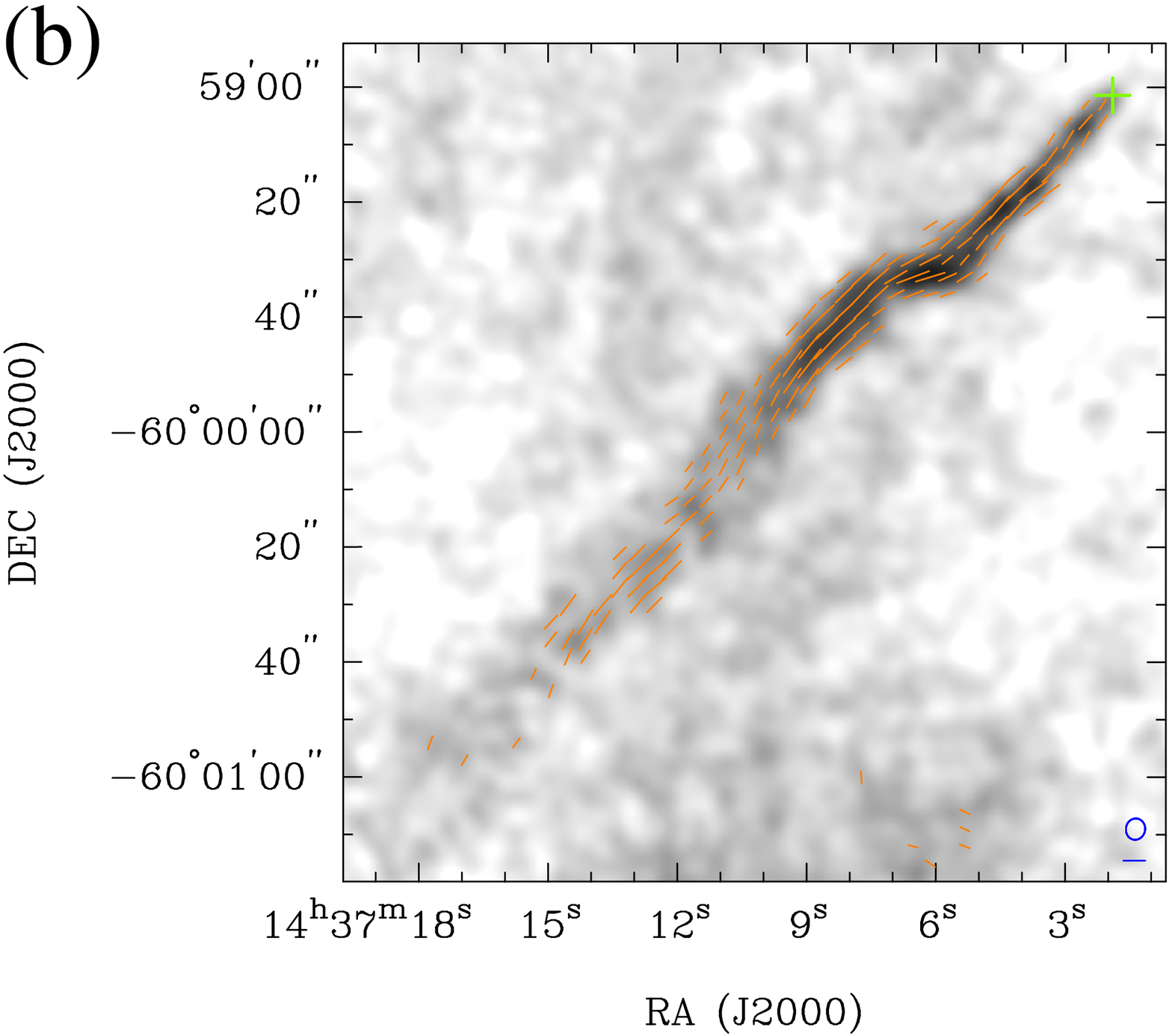}
\plottwo{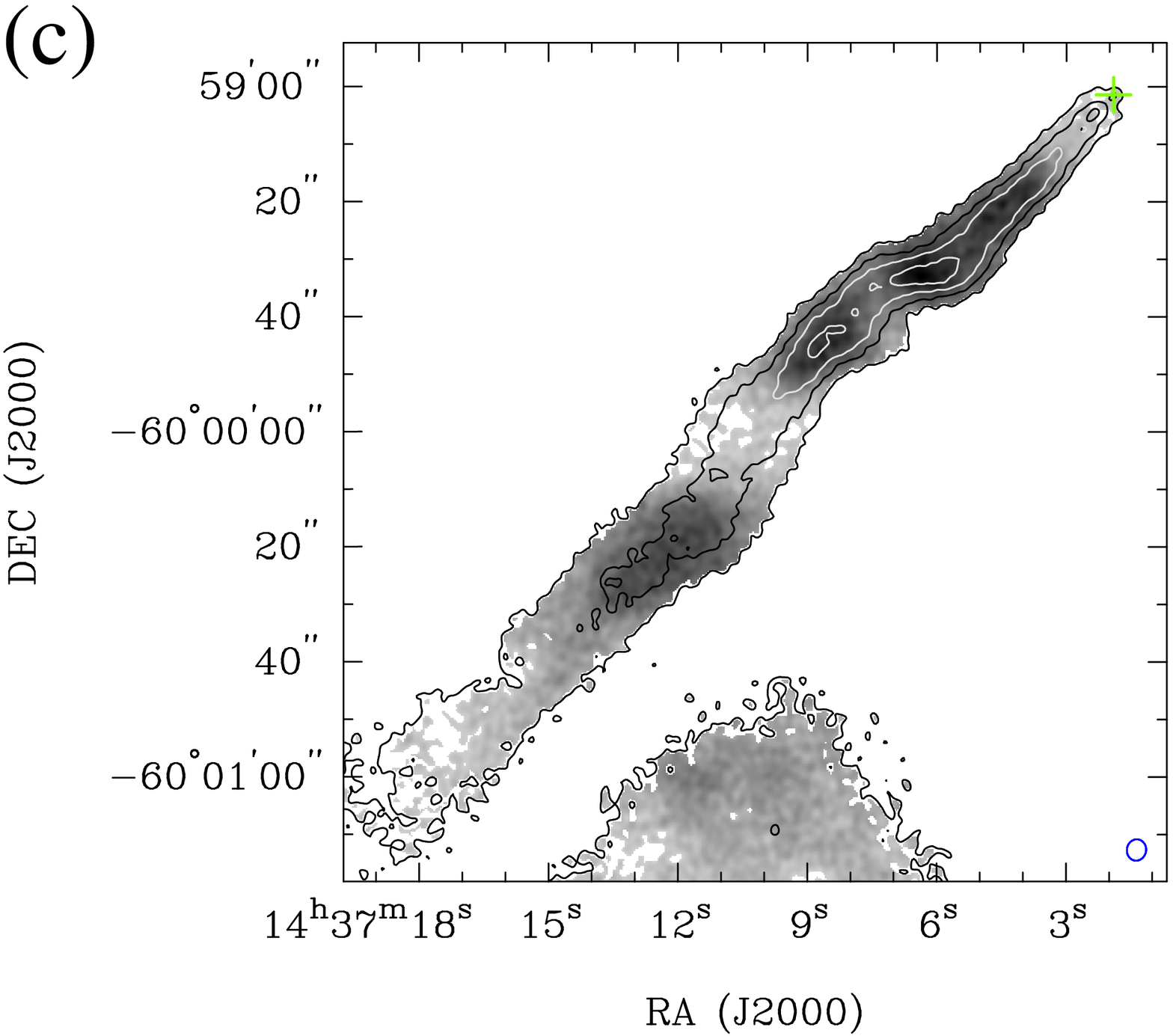}{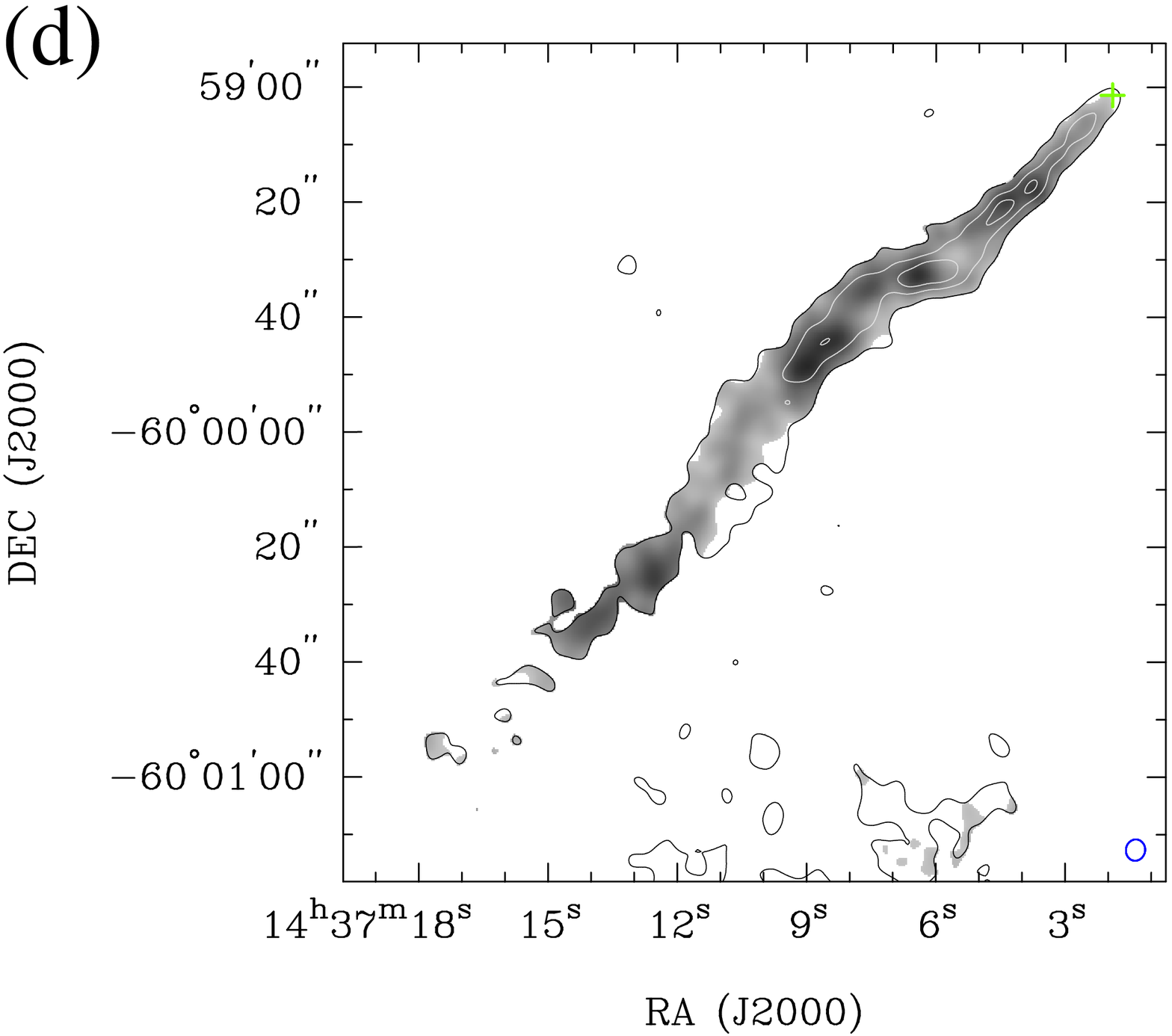}
\caption{Top: radio continuum images of the tip of \snr\ at: (a) 6\,cm
and (b) 3\,cm, overlaid with the corresponding polarization $B$-vectors that
show the intrinsic magnetic field orientation. The gray scales are linear,
ranging from $-0.045$ to 0.22\mjb\ at 6\,cm and from $-0.02$ to 0.3\mjb\ at
3\,cm. The vector lengths are proportional to the polarized intensity, with
the scale bars at the lower right representing 0.1\mjb. Bottom: linear
polarized emission at: (a) 6\,cm and (b) 3\,cm, overlaid with total intensity
contours at levels of 0.03, 0.073, 0.15, and 0.22\mjb\ for 6\,cm and 0.06,
0.12, and 0.17\mjb\ for 3\,cm. The gray scales are linear, ranging from 0 to
0.15\mjb\ at 6\,cm and from 0 to 0.2\mjb\ at 3\,cm. In all panels, position of
\psr\ is marked by the cross and the restoring beams are shown at the lower
right. \label{fig:pl}\vspace*{2mm}}
\end{figure*}

As shown in Figure~\ref{fig:img}, the handle is highly collimated and extends
8\farcm5 from the shell to the pulsar, at a position angle of 314\arcdeg\
(north through east). It has a width of 6\arcsec\ at the tip, which is only
resolved in the 3 and 6\,cm images, then gradually widens to $\sim60\arcsec$
at the intersection with the shell. At the source distance of $8\pm2$\,kpc,
these correspond to a physical length of $20\pm5$\,pc and width from
$0.23\pm0.06$\,pc to $2.3\pm0.6$\,pc. Beyond 5\arcmin\ from the pulsar, the
3\,cm emission falls below the observation sensitivity. We note that
extrapolating the handle inside the remnant slightly misses the shell's
geometric center by $\sim$1\farcm5 to the northeast.

The entire handle has flux densities of $110\pm40$, $35\pm1$, $26\pm10$ and
$18\pm5$\,mJy at 36, 13, 6, and 3\,cm, respectively. These measurements have
large uncertainties due to strong variation in the diffused background.
Therefore, we did not attempt to derive the radio spectra index from them,
instead, we employed the spectral tomography technique between the 3 and 6\,cm
data. We followed a similar procedure as described by \citet{gw03} and
smoothed the maps to 10\arcsec\ resolution to boost the signal-to-noise ratio.
We found a spectral index $\alpha=-0.45\pm0.1$ (defined as $S_\nu\propto
\nu^\alpha$), but no significant spatial variation of $\alpha$.

\subsection{The Handle}
Figures~\ref{fig:pl}(a) and (b) show the 3 and 6\,cm intensity maps
zoomed-in near the pulsar. The detailed morphology of the handle is very
similar at both wavelengths. There is a hint of unresolved compact emission at
the pulsar position at 6\,cm (see the contours in Figure~\ref{fig:pl}(c)) with a
flux density $60\pm30\mu$Jy. However, this is not observed at 3\,cm. The radio
emission of the handle peaks at 45\arcsec\ from the pulsar, with flux
densities of 0.25 and 0.20\mjb\ at 6 and 3\,cm, respectively. Further away,
there is a kink at 1\arcmin\ from the pulsar, such that the handle changes
orientation by 34\arcdeg\ and runs towards east for 10\arcsec, then switches
back to the original direction. The reverse occurs around 30\arcsec\ further
southeast. As a net result, a 40\arcsec-long section is displaced by 6\arcsec\
to the northeast.

For polarimetry results, the polarization $B$-vectors, after correcting for
foreground Faraday rotation, are overplotted in Figures~\ref{fig:pl}(a) and
(b). These indicate a highly ordered magnetic field structure. The
$B$-field well aligns with the handle's elongation, and changes direction
along with the handle across the kink. Farther from the kink, the field
orientation remains constant until the polarization signal falls below
detection beyond 2\farcm4 from the pulsar. Figures~\ref{fig:pl}(c) and
(d) show the polarized intensity maps at 6 and 3\,cm.
The polarized emission generally follows the total intensity, except across
the kink, where it shows a slight depolarization. We estimate that the
polarization fraction is up to $\sim$40\% and $\sim$50\% at 6 and 3\,cm,
respectively. A reduced polarization fraction at lower frequency has been
observed in other sources \citep[e.g.,][]{ngc+10} and it could be attributed
to beam depolarization. However, we note that these values are rough estimates
due to high background. Also, variations in polarization angle could shift
power in the polarized image into smaller scales, to which the interferometric
observations are more sensitive \citep[see discussion in][]{gbm+99}.
Therefore, the absolute polarized flux could be biased and the maps should be
taken as relative measurements only.

\section{DISCUSSION}

\subsection{Nature of the Handle and Its Association with the SNR}
Our results show that the handle of the Frying Pan (\pwn) has a
high degree of linear polarization with a flat radio spectrum, hence
confirming that it is a PWN powered by \psr. If the source is at a distance of
8\,kpc, then it would have a physical length over 20\,pc, the longest pulsar
tail ever detected. The source morphology suggests that it is a bow shock with
the pulsar traveling highly supersonically in the ISM. As radio PWNe generally
have a very long lifetime (see Equation~(\ref{eqt:cooling}) below), the fact
that the tail points toward the general direction of the remnant center
argues for an association between \psr\ and \snr\ \citep[but see,][]{zbc+08}.
We performed Monte Carlo simulations to quantify the probability of chance
association. We employed a realistic population synthesis model
(\citealt{fk06}; A.\ Bouchard et al.\ 2012, in preparation) and found that at
this sky position, an average of 0.2 field radio pulsars are expected within
20\arcmin\ from the shell, younger than $10^6$\,yr and with a radio flux
larger than that of \psr. If we further require the pulsar's projected motion
in the past to intercept any part of the shell, then the number greatly
reduces to 0.0012, therefore, strongly suggesting an association between \psr\
and the Frying Pan. Although the tail does not point toward the exact geometry
center of the shell, this is not uncommon and could be the result of
asymmetric supernova explosion, the progenitor's proper motion, or a slight
gradient in the ISM density \citep[see,][]{gva06,nrb+07,vmc+07}.

To estimate the kinematics of the system, we extrapolate the tail inside the
shell and take the closest approach to the center as the supernova site, which
is 14\farcm5 away from the pulsar location. This implies a distance
$l=1.0\times10^{20}\,d_8$\,cm traveled by the pulsar since birth, at the
source distance of $8\,d_8$\,kpc. Here we assume the pulsar motion lies in the
plane of the sky for simplicity. The pulsar velocity $v_\ast$ and age $t$ are
then related by $v_\ast=l/t$. Even if the pulsar travels as fast as 2000\kms,
the system would still be older than 15\,kyr, such that \snr\ has long passed
the free expansion phase. On the other hand, the transition from the Sedov
phase to the snowplow phase occurs at $t_{\rm tr}\approx 29 E_{51}^{4/17}
n_0^{-9/17}$\,kyr with a shell radius $R_{\rm tr}\approx 19.1E_{51}^{5/17}
n_0^{-7/17}$, where $E_{51}$ is the supernova explosion energy in units of
$10^{51}$\,erg and $n_0$ is the ISM number density in cm$^{-3}$
\citep{bwb+98}. Our observed shell radius of 7\arcmin\ gives a physical size
$R=16.3\,d_8$\,pc, slightly smaller than $R_{\rm tr}$ for the typical values
of $E_{51}=1$ and $n_0$ between 0.1 and 1. Therefore, we adopt the Sedov solution
\begin{equation}
R=1.15\left(\frac{E_0t^2}{\rho}\right)^{1/5}\ , \label{eqt:snr}
\end{equation} where $E_0=10^{51}E_{51}$\,erg and $\rho$ is
the ISM density in g\,cm$^{-3}$.

\subsection{Pulsar Velocity and Age}
The bow-shock stand-off distance $r_{\rm bs}$ ahead
of the PWN is determined by the ram pressure balance between the pulsar
wind and the ISM. For isotropic wind, 
\begin{equation}
\rho v_\ast^2 =\frac{\dot{E}}{4\pi c r_{\rm bs}^2} \ . \label{eqt:pwn}
\end{equation}
Assuming a constant $\rho$ across the field, we can solve Equations
(\ref{eqt:snr}) and (\ref{eqt:pwn}) to obtain
\begin{equation}
r_{\rm bs}=\Bigg(\frac{\dot{E}}{4\pi c E_0 l^2}\Bigg)^{1/2}
\Bigg(\frac{R}{1.15}\Bigg)^{5/2} 
=0.0024\,d_8^{3/2}E_{51}^{-1}\,\mathrm{pc}
\ .
\end{equation}
This corresponds to an angular scale $0\farcs06\,d_8^{1/2}\,E_{51}^{-1}$, too
small to be resolved by the observations. It is interesting to note that this
value is independent of the system age, pulsar velocity and ambient density.
Also, it is one of the smallest stand-off distances among bow-shock PWNe,
e.g., an order of magnitude smaller than that of the Mouse \citep{gvc+04}.
This indicates either a very large Mach number or a very dense ambient medium.
We argue that the latter case is less likely, since the CO map \citep{dht01}
shows no evidence of molecular clouds in the region. As mentioned, there is a
marginal detection of a compact feature at the tip of the PWN with a flux
density $60\pm30\mu$Jy at 6\,cm. If this is from the pulsar, then comparing to
the pulsed flux density of 75$\mu$\,Jy at 20\,cm \citep{cng+09} indicate a
rather flat radio spectrum. As an alternative, this could also be the surface
of the backward termination shock, which has an extent of $5r_{\rm bs}$
according to hydrodynamic simulations \citep{gvc+04}, such a small-scale
structure would not be resolved in our images. Further observations with high
time resolution could distinguish between these cases. Finally, we note that
the PWN half-width near the tip is about 40$r_{\rm bs}$, much greater than the
value found in magnetohydrodynamic (MHD) simulations \citep{bad05}. The
discrepancy could possibly be due to anisotropy of the pulsar wind or mass
loading through contamination \citep[see][]{vmc+07,rsc+10,lyu03}.

Assuming cosmic abundances, we substitute $r_{\rm bs}$ into
Equation~(\ref{eqt:pwn}) to obtain
\begin{equation}
v_\ast=1700\,E_{51}^{1/2}d_8^{-3/2}n_0^{-1/2}\,
\mathrm{km\,s}^{-1}
\end{equation}
and hence
\begin{equation}
t=19\,E_{51}^{-1/2}d_8^{5/2}n_0^{1/2}\,\mathrm{kyr} \ . \label{eqt:age}
\end{equation}
While we argue above that $n_0$ cannot be too high, $n_0$ cannot be too
low either because of the detection of the radio shell in \snr, hence, we
can rule out the hot ISM phase. For a typical value of $n_0$ between 0.1
and 1 in the warm ISM phase, the inferred velocity would be nearly
2000\kms with a young system age in the order of 10\,kyr, as proposed
by \citet{cng+09}. This indicates $t<t_{\rm tr}$, providing support to
the Sedov phase for the SNR. We also note that the age estimate is
much younger than the pulsar's characteristic age of 110\,kyr, which
could be reconciled if the pulsar was born with a spin period close to
the present-day value \citep[see][]{nrb+07}.
With these results, we can estimate the pressure
$P_{\rm head}$ at the head of the bow shock and the pulsar
Mach number $\mathcal{M}$ using
\begin{equation}
\gamma P_{\rm ISM} \mathcal{M}^2=P_{\rm head}=\rho v_\ast^2 =
6.8\times10^{-8}\,E_{51}d_8^{-3}\,\mathrm{dyn\,cm^{-2}} \ ,
\end{equation}
where $P_{\rm ISM}$ is the ambient pressure and $\gamma=5/3$ is the ISM
adiabatic index. Since different phases of the ISM are generally in pressure
equilibrium, $P_{\rm ISM}$ should be more uniform than $\rho$, hence, the Mach
number estimate would be more certain than the pulsar velocity estimate
\citep[see the detailed discussion by][]{kmp+08}. A typical ISM pressure of
$10^{-12} $\,dyn\,cm$^{-2}$ (corresponding to $P_{\rm ISM}/k=7250\,
\mathrm{cm^{-3}K}$) gives $\mathcal{M}=200\, E_{51}^{1/2}d_8^{-3/2}$,
suggesting a highly supersonic pulsar motion. For smaller values of $P_{\rm
ISM}$ \citep[e.g.,][]{fer01,cox05}, the inferred Mach number would be even
higher. While such a highly supersonic motion is expected to produce a very
sharp Mach cone with a small half-opening angle $\theta\sim \sin^{-1}(1/200)$,
the observed width of the tail increases from 6\arcsec\ to 60\arcsec\ over its
entire length of 7\arcmin, implying $\theta\sim \sin^{-1}(1/15)$. The
discrepancy could be the result of overpressure in the tail, such that 
\begin{eqnarray}
P_{\rm tail} &\sim &\rho (v_\ast \sin \theta)^2=P_{\rm head}\sin^2 \theta
 \nonumber \\ 
 &=& P_{\rm head}/15^2 = 0.004P_{\rm head} \ . \label{eqt:overpressure}
\end{eqnarray}
See Section \ref{sect:pressure} below for further discussion.

\subsection{The Kink}
Figure~\ref{fig:pl} shows a kink in the tail at 1\arcmin\ from the pulsar. We
argue that this is unlikely caused by flow instability or by a global ISM
pressure gradient, since the magnetic field is highly ordered across the kink
and the tail resumes its original orientation further downstream. Instead, it
could be the result of interstellar turbulence. Comparing the kink's
6\arcsec-lateral displacement to its 1\arcmin-separation from the pulsar , we
can deduce a turbulent velocity $v_t=(6\arcsec/ 1\arcmin)\,v_\ast= 0.1v_\ast$,
which is much higher than the typical sound speed of $\sim$10\kms\ in the warm
ISM phase, giving evidence of supersonic turbulence. As this PWN system has a
rather high flow speed (see Section \ref{sect:pressure} below), the postshock wind
could become compressible. The compression by the ISM turbulence would result
in enhanced synchrotron emissivity, which could naturally explain the
coincidence of the kink and the radio peak. Also, it would introduce disorder
in the magnetic field, reducing the polarized fraction near the kink as
observed.

\subsection{Magnetic Field Strength}
To estimate the magnetic field strength in the PWN, we define $k_m=U_B/U_p$ to
be the magnetization of the postshock wind, i.e., the ratio between the
magnetic and relativistic particle (electron+ion) energy densities
\citep[see][]{ptk+03}. MHD simulations show a relatively uniform magnetization
in the tail with $k_m$ of the typical order of 0.1 \citep{bad05}. Let $\eta$
be the ion to electron energy density ratio, we then have $U_p=(1+\eta)U_e$
and we can estimate the average magnetic field strength $B$ from standard
synchrotron theory 
\begin{equation}
\frac{U_B}{k_m (1+\eta)}=U_e=c_{12}B^{-3/2} L/V \ ,
\end{equation}
where $c_{12}$ is a constant that depends weakly on the spectral index and on
the lower and upper frequency limits of the emission \citep[see][]{pac70} and
$L/V$ is the synchrotron volume emissivity. Assuming a simple power-law
spectrum between $10^7$ and $10^{13}$\,Hz as in \citet{ngc+10}, our measured flux
density and spectral index yield 
\begin{equation}
B=\bigg[8\pi c_{12} k_m (1+\eta)\,\frac{L}{V}\bigg]^{2/7}=
60\,k_m^{2/7}\,(1+\eta)^{2/7}\,d_8^{-2/7}\,\mu\mathrm{G} \ .
\end{equation}
We note that this result is insensitive to the frequency limits, and to $k_m$
and $\eta$. For example, changing the upper limit to $10^{11}$\,Hz only lowers
$B$ by 13\%. If we ignore any ions (i.e., $\eta=0$) and take $k_m=0.1$, we
obtain $B=30\mu$G. Adopting $k_m=0.01$ yields $B=16\mu$G, while the
equipartition case ($k_m\approx1$) gives $B=60\mu$G. For emission at 6\,cm
($\nu\approx5$\,GHz), the synchrotron lifetime is
\begin{equation}
t_{\rm syn}=18B^{-3/2}\,\nu^{-1/2}\,\mathrm{kyr}=540\,k_m^{-3/7}(1+
\eta)^{-3/7}\,d_8^{3/7}\,\mathrm{kyr} \ , \label{eqt:cooling}
\end{equation}
sufficiently longer than the estimated age of the system (see Equation
(\ref{eqt:age})). Comparing to the synchrotron cooling, inverse Compton
scattering with the cosmic microwave background is negligible. Therefore, the
radio PWN should trace the pulsar motion over a long time, providing support
to the association with \snr.

\subsection{Pressure and Flow speed}\label{sect:pressure}
The magnetic field estimate can give us a handle on $P_{\rm ISM}$, which
consists of the magnetic pressure and the particle pressure. Taking the
pulsar wind as relativistic ideal gas, the particle pressure is given by
$U_p/3$, hence,
\begin{eqnarray}
P_{\rm tail}&=\frac{B^2}{8\pi}+\frac{U_p}{3}=
\big(1+\frac{1}{3k_m}\big)\frac{B^2}{8\pi}=1.5\times10^{-10}
 \nonumber \\
&\big(1+\frac{1}{3k_m}\big)k_m^{4/7}(1+\eta)^{4/7}
d_8^{-4/7}\,\mathrm{dyn\,cm^{-2}} \ .
\end{eqnarray}
This confirms that the system is highly overpressured as compared to the
environment. To match $P_{\rm tail}$ we derived in Equation
(\ref{eqt:overpressure}), it would require $\eta <1$ and $k_m\sim0.02$, hence,
this gives some hints that the postshock wind is dominated by electrons and
positrons with a small magnetization. Finally, we note that $P_{\rm tail}$
above is much lower than the value $0.02P_{\rm head}$ found in MHD simulations
\citep{bad05}, although those simulations only extend for a few $r_{\rm bs}$
in the tail direction.

We can estimate the average flow speed $v_f$ of the postshock wind based on
energy conservation. If we assume the pulsar spin-down energy all converts
into the magnetic and particle kinetic energy downstream, then
\begin{equation}
\dot{E} = A(v_f+v_\ast)(U_B+U_p)=A(v_f+v_\ast)
\Bigg(1+\frac{1}{k_m}\Bigg)\frac{B^2}{8\pi} \ ,
\end{equation}
where $A$ is the tail's cross-sectional area. We modeled the tail with a
truncated cone of diameter from 6\arcsec\ to 60\arcsec, then $A$ scales
with the distance $z$ (in arcminutes) from the pulsar as $A(z)\approx
4.0\times10^{35}(1+1.1z)^2d_8^2$\,cm$^{-2}$, giving
\begin{equation}
v_f \approx 2.3\times10^5\,(1+1.1z)^{-2}\frac{k_m^{3/7}}{1+k_m}
(1+\eta)^{-4/7}d_8^{-10/7} \,\mathrm{km\,s^{-1}} \ ,
\end{equation}
here $v_\ast$ is ignored as it is negligible. Our result indicates a rather
high flow speed near the pulsar ($z\lesssim1$): for $\eta=0$ and $k_m$=0.1, we
obtain $v_f\approx 0.06c$. If $k_m$=0.01, then $v_f$ is a factor of 2.5 lower.
These numbers are slightly higher than those observed in other
bow-shock PWN systems \citep[e.g.,][]{gvc+04,kmp+08}, but lower than the
simulation prediction \citep{bad05}. We note that while synchrotron loss is
negligible, other energy dissipation mechanisms, e.g., induced by flow
instability, mass-loading or ions in the wind, could significantly reduce the
flow speed \citep[see][]{lyu03}. Combining our radio results with future X-ray
observations can provide a more complete picture of the particle spectrum,
thus, better accounting for the total energetics of the wind to reveal the
flow structure. Spatially-resolved X-ray spectroscopy can also indicate any
deceleration in the flow. For particles emitting in the X-ray range ($\nu\sim
5\times10^{17}$\,Hz), Equation~(\ref{eqt:cooling}) gives a synchrotron cooling
time of 150\,yr. Hence, a flow speed of 0.06$c$ would result in a 3\,pc-long
X-ray PWN, comparable to other long X-ray tails \citep{kmp+08}.

\subsection{Comparison with Other Long Pulsar Tails in Radio}
In Table~\ref{table:pwn}, we summarize the physical extents and space
velocities of the longest pulsar tails detected in radio. We found only a very
week correlation between these two quantities. This is not surprisingly, since
we expect that the radio emission should also depend on the physical
conditions of the local environment, such as ambient density and magnetic
field. The three longest radio tails: the Frying Pan handle (20\,pc), the
Mouse (16\,pc) and G319.9$-$0.7 (10\,pc), all have very high Mach numbers
\citep[200, 60 and 30, respectively;][]{hgc+09,kmp+08} and exhibit smooth
morphologies with highly ordered magnetic fields. This indicates a smooth flow
structure in these systems, which helps explain the high degree of collimation
over long distances.

On the other hand, these three sources show some discrepancies in the detailed
morphology and magnetic field structure. The radio emission of the Mouse peaks
at the head then gradually fades downstream, and it shows a cone-shaped outer
region and a narrow inner region. For the handle of the Frying Pan, the tip is
very faint and a two-component structure is not obvious, although the
unresolved source at the pulsar position may correspond to the inner region,
or the outer region could be too faint to detect. Both the Frying Pan handle
and the Mouse have a very similar magnetic field geometry that runs parallel
to the tail \citep{yg05}, while G319.9$-$0.7 shows a $B$-field perpendicular
to the tail \citep{ngc+10}. The origin of this diversity is unclear. It may
depend on the pulsar spin orientation relative to its motion or the flow
condition of the system. In the regime of high postshock flow speed, the
magnetic field may be sheared by the flow so that the field lines are
stretched along the tail. More examples are needed to verify this picture.
Finally, we note that the polarization vectors in both the Mouse and
G319.9$-$0.7 exhibit abrupt switches at large distances from the pulsar
\citep{yb87,ngc+10}, which is not observed here. Therefore, we believe that
this is not a universal feature among long pulsar tail.

\section{CONCLUSIONS}

In this paper, we have presented a detailed radio study focusing on the handle
of the Frying Pan SNR (\snr) using ATCA and MOST observations. Our results
confirm the 20\,pc-long protrusion as a bow-shock PWN associated with \psr.
The detection of both the PWN and the parent SNR makes this system a powerful
probe of the pulsar environment. The very small bow-shock stand off distance
of 0.002\,pc we deduced implies a large pulsar velocity over 1000\kms\ and a
high Mach number $\sim$200. The PWN has a highly ordered magnetic field that
runs parallel to the tail orientation and has an average strength
$\sim$20$\mu$G. Based on the pressure estimate and the tail's opening angle,
we found some hints of a small magnetization in the postshock wind of the
order of 0.01--0.1. Also, the ions in the wind could be less energetic than
the electrons and positrons. Finally, we suggest that the kink observed in the
tail is a result of supersonic ISM turbulence.

As a caveat, the main uncertainties in our derived quantities come from the
distance estimate. It is not uncommon for dispersion measure-based distances
to be off by 25\% or more. While we may have to wait for the Square Kilometer
Array for a direct distance measurement with parallax \citep{stw+11},
multiwavelength studies are essential to further understand the physics of
this remarkable system. In particular, X-rays observations can provide a
complementary picture of the particle energetics and reveal the flow structure
of the highest energy particles. If TeV emission is detected, as in other
PWNe, a joint modeling of the synchrotron and inverse Compton scattering
emissions can give an accurate measurement of the magnetic field strength. For
a pulsar space velocity of 2000\kms, a direct proper motion measurement with
the ATCA would require a time span over 10\,yr. Therefore, other techniques,
such as vey long baseline interfermometry or timing observations, would be more practical. The latter is
also essential for a deeper search of $\gamma$-ray pulsations in future
\emph{Fermi} data.

\acknowledgements
We thank the referee for careful reading and useful comments. We thank Gemma
Anderson to help carry out the 2009 June ATCA observations.  The Australia
Telescope is funded by the Commonwealth of Australia for operation as a
National Facility managed by CSIRO. The MOST is operated by the University of
Sydney with support from the Australian Research Council and the Science
Foundation for Physics within The University of Sydney. 

{\it Facilities:} \facility{Molonglo Observatory ()}, \facility{ATCA ()}


\begin{thebibliography}{40}
\expandafter\ifx\csname natexlab\endcsname\relax\def\natexlab#1{#1}\fi

\bibitem[{{Becker} {et~al.}(2006){Becker}, {Kramer}, {Jessner}, {Taam}, {Jia},
  {Cheng}, {Mignani}, {Pellizzoni}, {de Luca}, {S{\l}owikowska}, \&
  {Caraveo}}]{bkj+06}
{Becker}, W., {et~al.} 2006, \apj, 645, 1421

\bibitem[{{Blondin} {et~al.}(1998){Blondin}, {Wright}, {Borkowski}, \&
  {Reynolds}}]{bwb+98}
{Blondin}, J.~M., {Wright}, E.~B., {Borkowski}, K.~J., \& {Reynolds}, S.~P.
  1998, \apj, 500, 342

\bibitem[{{Brentjens} \& {de Bruyn}(2005)}]{bd05}
{Brentjens}, M.~A., \& {de Bruyn}, A.~G. 2005, \aap, 441, 1217

\bibitem[{{Bucciantini} {et~al.}(2005){Bucciantini}, {Amato}, \& {Del
  Zanna}}]{bad05}
{Bucciantini}, N., {Amato}, E., \& {Del Zanna}, L. 2005, \aap, 434, 189

\bibitem[{{Camilo} {et~al.}(2009){Camilo}, {Ng}, {Gaensler}, {Ransom},
  {Chatterjee}, {Reynolds}, \& {Sarkissian}}]{cng+09}
{Camilo}, F., {Ng}, C.-Y., {Gaensler}, B.~M., {Ransom}, S.~M., {Chatterjee},
  S., {Reynolds}, J., \& {Sarkissian}, J. 2009, \apjl, 703, L55

\bibitem[{{Castelletti} {et~al.}(2003){Castelletti}, {Dubner}, {Golap}, {Goss},
  {Vel{\'a}zquez}, {Holdaway}, \& {Rao}}]{cdg+03}
{Castelletti}, G., {Dubner}, G., {Golap}, K., {Goss}, W.~M., {Vel{\'a}zquez},
  P.~F., {Holdaway}, M., \& {Rao}, A.~P. 2003, \aj, 126, 2114

\bibitem[{{Chang} {et~al.}(2011){Chang}, {Pavlov}, {Kargaltsev}, \&
  {Shibanov}}]{cpk+11}
{Chang}, C., {Pavlov}, G.~G., {Kargaltsev}, O., \& {Shibanov}, Y.~A. 2011, ApJ,
  in press, arXiv:1107.1819

\bibitem[{{Chatterjee} {et~al.}(2004){Chatterjee}, {Cordes}, {Vlemmings},
  {Arzoumanian}, {Goss}, \& {Lazio}}]{ccv+04}
{Chatterjee}, S., {Cordes}, J.~M., {Vlemmings}, W.~H.~T., {Arzoumanian}, Z.,
  {Goss}, W.~M., \& {Lazio}, T.~J.~W. 2004, \apj, 604, 339

\bibitem[{{Chen} {et~al.}(2006){Chen}, {Wang}, {Gotthelf}, {Jiang}, {Chu}, \&
  {Gruendl}}]{cwg+06}
{Chen}, Y., {Wang}, Q.~D., {Gotthelf}, E.~V., {Jiang}, B., {Chu}, Y.-H., \&
  {Gruendl}, R. 2006, \apj, 651, 237

\bibitem[{{Cox}(2005)}]{cox05}
{Cox}, D.~P. 2005, \araa, 43, 337

\bibitem[{{Dame} {et~al.}(2001){Dame}, {Hartmann}, \& {Thaddeus}}]{dht01}
{Dame}, T.~M., {Hartmann}, D., \& {Thaddeus}, P. 2001, \apj, 547, 792

\bibitem[{{Faucher-Gigu{\`e}re} \& {Kaspi}(2006)}]{fk06}
{Faucher-Gigu{\`e}re}, C.-A., \& {Kaspi}, V.~M. 2006, \apj, 643, 332

\bibitem[{{Ferri{\`e}re}(2001)}]{fer01}
{Ferri{\`e}re}, K.~M. 2001, Reviews of Modern Physics, 73, 1031

\bibitem[{{Frail} {et~al.}(1996){Frail}, {Giacani}, {Goss}, \&
  {Dubner}}]{fgg+96}
{Frail}, D.~A., {Giacani}, E.~B., {Goss}, W.~M., \& {Dubner}, G. 1996, \apjl,
  464, L165

\bibitem[{{Gaensler} {et~al.}(1999){Gaensler}, {Brazier}, {Manchester},
  {Johnston}, \& {Green}}]{gbm+99}
{Gaensler}, B.~M., {Brazier}, K.~T.~S., {Manchester}, R.~N., {Johnston}, S., \&
  {Green}, A.~J. 1999, \mnras, 305, 724

\bibitem[{{Gaensler} {et~al.}(2006){Gaensler}, {Chatterjee}, {Slane}, {van der
  Swaluw}, {Camilo}, \& {Hughes}}]{gcs+06}
{Gaensler}, B.~M., {Chatterjee}, S., {Slane}, P.~O., {van der Swaluw}, E.,
  {Camilo}, F., \& {Hughes}, J.~P. 2006, \apj, 648, 1037

\bibitem[{{Gaensler} \& {Slane}(2006)}]{gs06}
{Gaensler}, B.~M., \& {Slane}, P.~O. 2006, \araa, 44, 17

\bibitem[{{Gaensler} {et~al.}(2004){Gaensler}, {van der Swaluw}, {Camilo},
  {Kaspi}, {Baganoff}, {Yusef-Zadeh}, \& {Manchester}}]{gvc+04}
{Gaensler}, B.~M., {van der Swaluw}, E., {Camilo}, F., {Kaspi}, V.~M.,
  {Baganoff}, F.~K., {Yusef-Zadeh}, F., \& {Manchester}, R.~N. 2004, \apj, 616,
  383

\bibitem[{{Gaensler} \& {Wallace}(2003)}]{gw03}
{Gaensler}, B.~M., \& {Wallace}, B.~J. 2003, \apj, 594, 326

\bibitem[{{Gvaramadze}(2006)}]{gva06}
{Gvaramadze}, V.~V. 2006, \aap, 454, 239

\bibitem[{{Hales} {et~al.}(2009){Hales}, {Gaensler}, {Chatterjee}, {van der
  Swaluw}, \& {Camilo}}]{hgc+09}
{Hales}, C.~A., {Gaensler}, B.~M., {Chatterjee}, S., {van der Swaluw}, E., \&
  {Camilo}, F. 2009, \apj, 706, 1316

\bibitem[{{Haverkorn} {et~al.}(2006){Haverkorn}, {Gaensler},
  {McClure-Griffiths}, {Dickey}, \& {Green}}]{hgm+06}
{Haverkorn}, M., {Gaensler}, B.~M., {McClure-Griffiths}, N.~M., {Dickey},
  J.~M., \& {Green}, A.~J. 2006, \apjs, 167, 230

\bibitem[{{Kargaltsev} {et~al.}(2008){Kargaltsev}, {Misanovic}, {Pavlov},
  {Wong}, \& {Garmire}}]{kmp+08}
{Kargaltsev}, O., {Misanovic}, Z., {Pavlov}, G.~G., {Wong}, J.~A., \&
  {Garmire}, G.~P. 2008, \apj, 684, 542

\bibitem[{{Lazendic} {et~al.}(2000){Lazendic}, {Dickel}, {Haynes}, {Jones}, \&
  {White}}]{ldh+00}
{Lazendic}, J.~S., {Dickel}, J.~R., {Haynes}, R.~F., {Jones}, P.~A., \&
  {White}, G.~L. 2000, \apj, 540, 808

\bibitem[{{Lyutikov}(2003)}]{lyu03}
{Lyutikov}, M. 2003, \mnras, 339, 623

\bibitem[{{Ng} {et~al.}(2010){Ng}, {Gaensler}, {Chatterjee}, \&
  {Johnston}}]{ngc+10}
{Ng}, C.-Y., {Gaensler}, B.~M., {Chatterjee}, S., \& {Johnston}, S. 2010, \apj,
  712, 596

\bibitem[{{Ng} {et~al.}(2007){Ng}, {Romani}, {Brisken}, {Chatterjee}, \&
  {Kramer}}]{nrb+07}
{Ng}, C.-Y., {Romani}, R.~W., {Brisken}, W.~F., {Chatterjee}, S., \& {Kramer},
  M. 2007, \apj, 654, 487

\bibitem[{{Olbert} {et~al.}(2001){Olbert}, {Clearfield}, {Williams}, {Keohane},
  \& {Frail}}]{ocw+01}
{Olbert}, C.~M., {Clearfield}, C.~R., {Williams}, N.~E., {Keohane}, J.~W., \&
  {Frail}, D.~A. 2001, \apjl, 554, L205

\bibitem[{{Pacholczyk}(1970)}]{pac70}
{Pacholczyk}, A.~G. 1970, {Radio Astrophysics. Nonthermal Processes in Galactic
  and Extragalactic Sources} (San Francisco: Freeman)

\bibitem[{{Pavlov} {et~al.}(2003){Pavlov}, {Teter}, {Kargaltsev}, \&
  {Sanwal}}]{ptk+03}
{Pavlov}, G.~G., {Teter}, M.~A., {Kargaltsev}, O., \& {Sanwal}, D. 2003, \apj,
  591, 1157

\bibitem[{{Romani} {et~al.}(2010){Romani}, {Shaw}, {Camilo}, {Cotter}, \&
  {Sivakoff}}]{rsc+10}
{Romani}, R.~W., {Shaw}, M.~S., {Camilo}, F., {Cotter}, G., \& {Sivakoff},
  G.~R. 2010, \apj, 724, 908

\bibitem[{{Sault} {et~al.}(1999){Sault}, {Bock}, \& {Duncan}}]{sbd99}
{Sault}, R.~J., {Bock}, D.~C.-J., \& {Duncan}, A.~R. 1999, \aaps, 139, 387

\bibitem[{{Smits} {et~al.}(2011){Smits}, {Tingay}, {Wex}, {Kramer}, \&
  {Stappers}}]{stw+11}
{Smits}, R., {Tingay}, S.~J., {Wex}, N., {Kramer}, M., \& {Stappers}, B. 2011,
  \aap, 528, A108

\bibitem[{{van der Swaluw}(2004)}]{van04}
{van der Swaluw}, E. 2004, Advances in Space Research, 33, 475

\bibitem[{{Vigelius} {et~al.}(2007){Vigelius}, {Melatos}, {Chatterjee},
  {Gaensler}, \& {Ghavamian}}]{vmc+07}
{Vigelius}, M., {Melatos}, A., {Chatterjee}, S., {Gaensler}, B.~M., \&
  {Ghavamian}, P. 2007, \mnras, 374, 793

\bibitem[{{Whiteoak} \& {Green}(1996)}]{wg96}
{Whiteoak}, J.~B.~Z., \& {Green}, A.~J. 1996, \aaps, 118, 329

\bibitem[{{Wilson} {et~al.}(2011){Wilson}, {Ferris}, {Axtens}, {Brown},
  {Davis}, {Hampson}, {Leach}, {Roberts}, {Saunders}, {Koribalski}, {Caswell},
  {Lenc}, {Stevens}, {Voronkov}, {Wieringa}, {Brooks}, {Edwards}, {Ekers},
  {Emonts}, {Hindson}, {Johnston}, {Maddison}, {Mahony}, {Malu}, {Massardi},
  {Mao}, {McConnell}, {Norris}, {Schnitzeler}, {Subrahmanyan}, {Urquhart},
  {Thompson}, \& {Wark}}]{wfa+11}
{Wilson}, W.~E., {et~al.} 2011, \mnras, 416, 832

\bibitem[{{Yusef-Zadeh} \& {Bally}(1987)}]{yb87}
{Yusef-Zadeh}, F., \& {Bally}, J. 1987, \nat, 330, 455

\bibitem[{{Yusef-Zadeh} \& {Gaensler}(2005)}]{yg05}
{Yusef-Zadeh}, F., \& {Gaensler}, B.~M. 2005, Advances in Space Research, 35,
  1129

\bibitem[{{Zeiger} {et~al.}(2008){Zeiger}, {Brisken}, {Chatterjee}, \&
  {Goss}}]{zbc+08}
{Zeiger}, B.~R., {Brisken}, W.~F., {Chatterjee}, S., \& {Goss}, W.~M. 2008,
  \apj, 674, 271

\end{thebibliography}

\end{document}